\documentclass[]{piparticle-final}
\usepackage{graphicx}
\usepackage{amssymb}
\usepackage{amsmath}
\usepackage{cite}

\begin{document}
\volume{3}               
\articlenumber{030001}   
\journalyear{2011}       
\editor{G. C. Barker}   
\reviewers{B. Blasius, ICBM, University of Oldenburg, Germany.}  
\received{12 August 2010}     
\accepted{18 February 2011}   
\runningauthor{D. H. Zanette}  
\doi{030001}         

\title{SIR epidemics in monogamous populations with recombination}

\author{Dami\'an H. Zanette\cite{inst1}\thanks{E-mail: zanette@cab.cnea.gov.ar}}

\pipabstract{We study the propagation of an SIR
(susceptible--infectious--recovered) disease over an agent
population which, at any instant, is fully divided into couples of
agents. Couples are occasionally allowed to exchange their members.
This process of couple recombination can compensate the
instantaneous disconnection of the interaction pattern and thus
allow for the propagation of the infection. We study the incidence
of the disease as a function of its infectivity and of the
recombination rate of couples, thus characterizing the interplay
between the epidemic dynamics and the evolution of the population's
interaction pattern.}

\maketitle

\blfootnote{
\begin{theaffiliation}{99}
\institution{inst1} Consejo Nacional de Investigaciones
Cient\'{\i}ficas y T\'ecnicas, Centro At\'omico Bariloche e
Instituto Balseiro, 8400 Bariloche, R\'{\i}o Negro, Argentina.
\end{theaffiliation}
}

\section{Introduction}

Models of disease propagation are widely used to provide a stylized
picture of the basic mechanisms at work during epidemic outbreaks
and infection spreading \cite{anderson}. Within interdisciplinary
physics, they have the additional interest of being closely related
to the mathematical representation of such diverse phenomena as fire
propagation, signal transmission in neuronal axons, and  oscillatory
chemical reactions \cite{selfo}. Because this kind of model
describes the joint dynamics of large populations of interacting
active elements or agents, its most interesting outcome is the
emergence of self-organization. The appearance of endemic states,
with a stable finite portion of the population actively transmitting
an infection, is a typical form of self-organization in
epidemiological models \cite{murray}.

Occurrence of self-organized collective behavior has, however, the
{\it sine qua non} condition that information about the individual
state of agents must be exchanged between each other. In turn, this
requires the interaction pattern between agents not to be disconnected. Fulfilment of such requirement is usually assumed to
be granted. However, it is not difficult to think of simple
scenarios where it is not guaranteed. In the specific context of
epidemics, for instance, a sexually transmitted infection never
propagates in a population where sexual partnership is confined
within stable couples or small groups \cite{sex1}.

In this paper, we consider an SIR
(susceptible--infectious--recovered) epidemiological model
\cite{murray} in a monogamous population where, at any instant, each
agent has exactly one partner or neighbor \cite{sex1,sex2}. The
population is thus divided into couples, and is therefore highly
disconnected. However, couples can occasionally break up and their
members can then be exchanged with those of other broken couples. As
was recently demonstrated for SIS models \cite{bouzat,vazquez}, this
process of couple recombination can compensate to a certain extent
the instantaneous lack of connectivity of the population's
interaction pattern, and possibly allow for the propagation of the
otherwise confined disease. Our main aim here is to characterize
this interplay between recombination and propagation for SIR
epidemics.

In the next section, we review the SIR model and its mean field
dynamics. Analytical results are then provided for recombining
monogamous populations in the limits of zero and infinitely large
recombination rate, while the case of intermediate rates is studied
numerically. Attention is focused on the disease incidence --namely,
the portion of the population that has been infectious sometime
during the epidemic process-- and its dependence on the disease
infectivity and the recombination rates, as well as on the initial
number of infectious agents. Our results are inscribed in the
broader context of epidemics propagation on populations with
evolving interaction patterns
\cite{sex1,sex2,gross,gross2,risau1,risau2}.

\section{SIR dynamics and mean field description}

In the SIR model, a disease propagates over a population each of
whose members can be, at any given time, in one of three
epidemiological states: susceptible (S), infectious (I), or
recovered (R). Susceptible agents become infectious by contagion
from infectious neighbors, with probability $\lambda$ per neighbor
per time unit. Infectious agents, in turn, become recovered
spontaneously, with probability $\gamma$ per time unit. The disease
process S $\to$ I $\to$ R ends there, since recovered agents cannot
be infected again \cite{murray}.

With a given initial fraction of S and I--agents, the disease first
propagates by contagion but later declines due to recovery. The
population ends in an absorbing state where the infection has
disappeared, and each agent is either recovered or still
susceptible. In this respect, SIR epidemics differs from the SIS and
SIRS models, where --due to the cyclic nature of the disease,-- the
infection can asymptotically reach an endemic state, with a constant
fraction of infectious agents permanently present in the population.

Another distinctive equilibrium property of SIR epidemics is that
the final state depends on the initial condition. In other words,
the SIR model possesses infinitely many equilibria parameterized by
the initial states.

In a mean field description, it is assumed that each agent is
exposed to the average epidemiological state of the whole
population. Calling $x$ and $y$ the respective fractions of S and
I--agents, the mean field evolution of the disease is governed by
the equations
\begin{equation}
\begin{array}{lll} \label{mf}
\dot x &=& -k\lambda xy , \\
\dot y &=& k\lambda xy - y,
\end{array}
\end{equation}
where $k$ is the average number of neighbors per agent. Since the
population is assumed to remain constant in size, the fraction of
R--agents is $z=1-x-y$.  In the second equation of Eqs. (\ref{mf}), we have
assigned the recovery frequency the value $\gamma =1$, thus fixing
the time unit equal to $\gamma^{-1}$, the average duration of the
infectious state. The contagion frequency $\lambda$ is accordingly
normalized: $\lambda/\gamma \to \lambda$. This choice for $\gamma$
will be maintained throughout the remaining of the paper.

\begin{figure}[th]
\begin{center}
\includegraphics[width=0.45\textwidth]{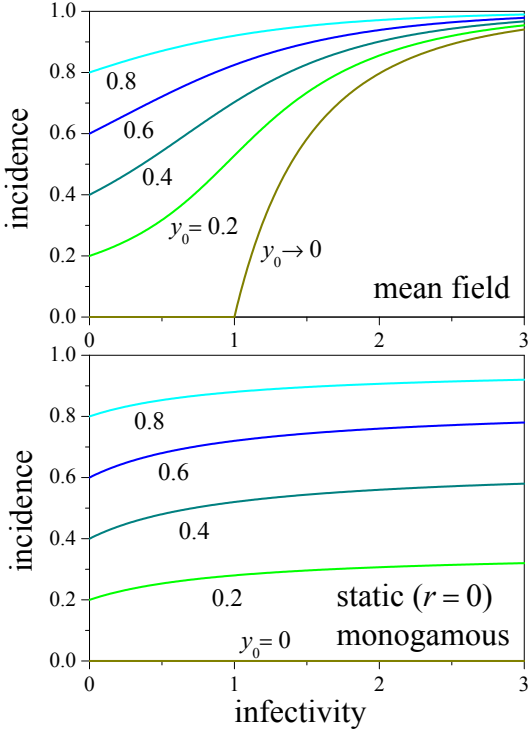}
\end{center}
\caption{SIR epidemics incidence (measured by the final fraction of
recovered agents $z^*$) as a function of the infectivity (measured
by the product of the mean number of neighbors times the infection
probability per time unit per infected neighbor, $k\lambda$), for
different initial fractions of infectious agents, $y_0$. Upper
panel: For the mean field equations (\ref{mf}). Lower panel: For a
static (non-recombining) monogamous population, described by Eqs.
(\ref{couples}) with $r=0$. } \label{fig1}
\end{figure}

The solution to Eqs. (\ref{mf}) implies that, from an initial
condition without R--agents, the final fraction of S--agents, $x^*$,
is related to the initial fraction of I--agents, $y_0$, as
\cite{anderson}
\begin{equation} \label{inc}
x^* = 1 - ( k \lambda)^{-1} \log [ (1-y_0)/ x^*] .
\end{equation}
Note that the final fraction of R--agents, $z^* = 1 - x^*$, gives
the total fraction of agents who have been infectious sometime
during the epidemic process. Thus, $z^*$ directly measures the total
 incidence of the disease.

The incidence $z^*$ as a function of the infectivity $k \lambda$,
obtained from Eq. (\ref{inc}) through the standard Newton--Raphson
method for several values $y_0$ of the initial fraction of
I--agents, is shown in the upper panel of Fig. \ref{fig1}. As
expected, the disease incidence grows both with the infectivity and
with $y_0$. Note that, on the one hand, this growth is smooth for
finite positive $y_0$. On the other hand, for $y_0\to 0$ (but
$y_0\ne 0$) there is a transcritical bifurcation at $k \lambda =1$.
For lower infectivities, the disease is not able to propagate and,
consequently, its incidence is identically equal to zero. For larger
infectivities, even when the initial fraction of I--agents is
vanishingly small, the disease propagates and the incidence turns
out to be positive. Finally, for $y_0=0$ no agents are initially
infectious, no infection spreads, and the incidence thus vanishes
all over parameter space.

\section{Monogamous populations with couple recombination}

Suppose now that, at any given time, each agent in the population
has exactly just one neighbor or, in other words, that the whole
population is always divided into couples. In reference to sexually
transmitted diseases, this pattern of contacts between agents
defines a {\it monogamous} population \cite{sex2}. If each couple is
everlasting, so that neighbors do not change with time, the disease
incidence should be heavily limited by the impossibility of
propagating too far from the initially infectious agents. At most,
some of the initially susceptible agents with infectious neighbors
will become themselves infectious, but spontaneous recovery will
soon prevail and the disease will disappear.

If, on the other hand, the population remains monogamous but
neighbors are occasionally allowed to change, any I--agent may
transmit the disease several times before recovering. If such
changes are frequent enough, the disease could perhaps reach an
incidence similar to that predicted by the mean field description,
Eq. (\ref{mf}) (for $k=1$, i.e. with an average of one neighbor per
agent).

We model neighbor changes by a process of couple recombination
where, at each event, two couples $(i,j)$ and $(m,n)$ are chosen at
random and their partners are exchanged \cite{bouzat,vazquez}. The
two possible outcomes of recombination, either $(i,m)$ and $(j,n)$
or $(i,n)$ and $(j,m)$, occur with equal probability. To quantify
recombination, we define  $r$ as  the probability per unit time that
any given couple becomes involved in such an event.

A suitable description of SIR epidemics in monogamous populations
with recombination is achieved in terms of the fractions of couples
of different kinds, $m_{\rm SS}$, $m_{\rm SI}$, $m_{\rm II}$,
$m_{\rm IR}$, $m_{\rm RR}$, and $m_{\rm SR}= 1-m_{\rm SI}-m_{\rm
II}-m_{\rm IR}-m_{\rm RR}$. Evolution equations for these fractions
are obtained by considering the possible transitions between kinds
of couples due to recombination and epidemic events \cite{vazquez}.
For instance, partner exchange between two couples (S,S) and (I,R)
which gives rise to (S,I) and (S,R), contributes positive terms to
the time derivative of $m_{\rm SI}$ and $m_{\rm SR}$, and negative
terms to those of $m_{\rm SS}$ and $m_{\rm IR}$, all of them
proportional to the product $m_{\rm SS} m_{\rm IR}$. Meanwhile, for
example, contagion can transform an (S,I)--couple into an
(I,I)--couple, with negative and positive contributions to the
variations of the respective fractions, both proportional to $m_{\rm
SI}$.

The equations resulting from these arguments read
\begin{equation} \label{couples}
\begin{array}{lll}
\dot m_{\rm SS} &=& r A_{\rm SIR}, \\
\dot m_{\rm SI} &=& r B_{\rm SIR}-(1+\lambda) m_{\rm SI} ,\\
\dot m_{\rm II} &=& r A_{\rm IRS}  +\lambda m_{\rm SI}- 2  m_{\rm II} ,\\
\dot m_{\rm IR} &=& r B_{\rm IRS} + 2  m_{\rm II}-  m_{\rm IR},\\
\dot m_{\rm RR} &=& r A_{\rm RSI} + m_{\rm IR}, \\
\dot m_{\rm SR} &=& r B_{\rm RSI} + m_{\rm SI}.
\end{array}
\end{equation}
For brevity, we have here denoted the contribution of recombination
by means of the symbols
\begin{equation}
A_{ijh}\equiv ( m_{ij}+m_{ih} )^2 /4 - m_{ii}  (
m_{jj}+m_{jh}+m_{hh} ),
\end{equation}
and
\begin{eqnarray}
&B_{ijh} \equiv   ( 2 m_{ii}+m_{ih}  ) ( 2 m_{jj}+m_{jh}  )/2\nonumber \\
& \ \ \ \ \ -  m_{ij}  ( m_{ij}+m_{ih}+m_{jh} + m_{hh} )/2,
\end{eqnarray}
with $i$, $j$, $h$ $\in \{ {\rm S, I, R}\}$. The remaining terms
stand for the epidemic events. In terms of the couple fractions, the
fractions of S, I and R--agents are expressed as
\begin{equation} \label{relat}
\begin{array}{lll}
x &=& m_{\rm SS} +   ( m_{\rm SI} +m_{\rm SR}  )/2, \\
y &=& m_{\rm II} +   ( m_{\rm SI} +m_{\rm IR}  )/2, \\
z &=& m_{\rm RR} +   ( m_{\rm SR} +m_{\rm IR}  )/2. \\
\end{array}
\end{equation}
Assuming that the agents with different epidemiological states are
initially distributed at random over the pattern of couples, the
initial fraction of each kind of couple is $m_{\rm SS} (0) = x_0^2$,
$m_{\rm SI}(0) = 2x_0 y_0$, $m_{\rm II}(0) = y_0^2$, $m_{\rm IR}
(0)= 2y_0 z_0$, $m_{\rm RR} (0)= z_0^2$, and $m_{\rm SR} (0)= 2x_0
z_0$, where $x_0$, $y_0$ and $z_0$ are the initial fractions of each
kind of agent.

It is important to realize that the mean field--like Eqs.
(\ref{couples}) to (\ref{relat}) are {\it exact} for infinitely
large populations. In fact, first, pairs of couples are selected at
random for recombination. Second, any epidemic event that changes
the state of an agent modifies the kind of the corresponding couple,
but does not affect any other couple. Therefore, no correlations are
created by either process.

In the limit without recombination, $r=0$, the pattern of couples is
static. Equations (\ref{couples}) become linear and can be
analytically solved. For asymptotically long times, the solution
provides --from the third of Eqs. (\ref{relat})-- the disease
incidence as a function of the initial condition. If no R--agents
are present in the initial state, the incidence is
\begin{equation}
z^* =  (1+\lambda)^{-1} [1+\lambda (2-y_0)] y_0.
\end{equation}
This is plotted in the lower panel of Fig. \ref{fig1} as a function
of the infectivity $k \lambda \equiv \lambda$, for various values of
the initial fraction of I--agents, $y_0$. When recombination is
suppressed, as expected, the incidence is limited even for large
infectivities, since disease propagation can only occur to
susceptible agents initially connected to infectious neighbors.
Comparison with the upper panel makes apparent substantial
quantitative differences with the mean field description, especially
for small initial fractions of I--agents.

Another situation that can be treated analytically is the limit of
infinitely frequent recombination, $r\to \infty$. In this limit,
over a sufficiently short time interval, the epidemiological state
of all agents is virtually ``frozen'' while the pattern of couples
tests all possible combinations of agent pairs. Consequently, at
each moment, the fraction of couples of each kind is completely
determined by the instantaneous fraction of each kind of agent,
namely,
\begin{equation} \label{relat2}
\begin{array}{ll}
&m_{\rm SS}   = x^2, \ \ m_{\rm SI} = 2x  y, \ \  m_{\rm II} = y^2, \\
&m_{\rm IR} = 2y z , \ \  m_{\rm RR}  = z ^2 , \ \  m_{\rm SR}  = 2x
z .
\end{array}
\end{equation}
These relations are, of course, the same  as quoted above for
uncorrelated initial conditions.

Replacing Eqs.  (\ref{relat2}) into (\ref{couples}) we verify,
first, that the operators $A_{ijh}$ and $B_{ijh}$ vanish
identically. The remaining of the equations, corresponding to the
contribution of epidemic events, become equivalent to the mean field
equations (\ref{mf}). Therefore, if the distributions of couples and
epidemiological states are initially uncorrelated, the evolution of
the fraction of couples of each kind is exactly determined by the
mean field description for the fraction each kind of agent, through
the relations given in Eqs. (\ref{relat2}).

For intermediate values of the recombination rate, $0<r<\infty$, we
expect to obtain incidence levels that interpolate between the
results presented in the two panels of Fig. \ref{fig1}. However,
these cannot be obtained analytically. We thus resort to the
numerical solution of Eqs. (\ref{couples}).

\begin{figure}[th]
\begin{center}
\includegraphics[width=0.48\textwidth]{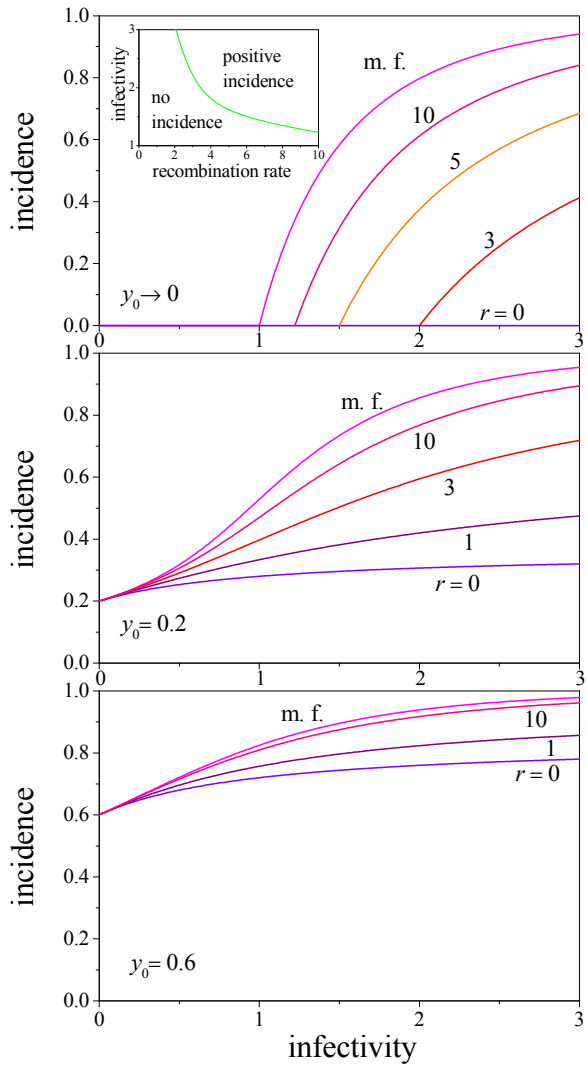}
\end{center}
\caption{SIR epidemics incidence  as a function of the infectivity
for three initial fractions of infectious agents, $y_0$, and several
recombination rates, $r$. Mean field (m.~f.) results are also shown.
The insert in the upper panel displays the boundary between the
phases of no incidence and positive incidence for $y_0 \to 0$, in
the parameter plane of infectivity vs. recombination rate. }
\label{fig2}
\end{figure}

\section{Numerical results for recombining couples}

We solve Eqs. (\ref{couples}) by means of a standard fourth-order
Runge-Kutta algorithm. The initial conditions are as in the
preceding section, representing no R--agents and a fraction $y_0$ of
I--agents. The disease incidence $z^*$ is estimated from the third equation 
of Eqs. (\ref{relat}), using the long-time numerical solutions for
$m_{\rm RR}$,   $m_{\rm SR}$, and  $m_{\rm IR}$. In the range of
parameters considered here, numerical integration up to time
$t=1000$ was enough to get a satisfactory approach to asymptotic
values.

Figure \ref{fig2} shows the incidence as a function of infectivity
for three values of the initial fraction of I--agents, $y_0 \to 0$,
$y_0=0.2$ and $0.6$, and several values of the recombination rate
$r$. Numerically, the limit $y_0 \to 0$ has been represented by
taking $y_0 = 10^{-9}$. Within the plot resolution, smaller values
of $y_0$ give identical results. Mean field (m.~f.) results are also
shown. As expected from the analytical results presented in the
preceding section, positive values of $r$ give rise to incidences
between those obtained for a static couple pattern ($r=0$) and for
the mean field description. Note that substantial departure from the
limit of static couples is only got for relatively large
recombination rates, $r > 1$, when at least one recombination per
couple occurs in the typical time of recovery from the infection.

\begin{figure}[th]
\begin{center}
\includegraphics[width=0.48\textwidth]{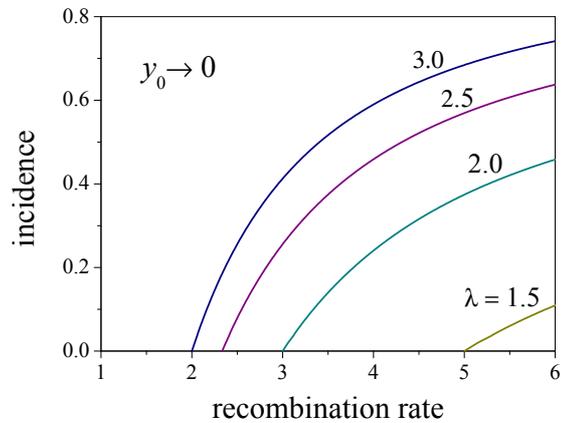}
\end{center}
\caption{SIR epidemics incidence as a function of the recombination
rate $r$ for a vanishingly small fraction of infectious agents,
$y_0\to 0$, and several infectivities $\lambda$.}  \label{fig3}
\end{figure}

Among these results, the most interesting situation is that of a
vanishingly small initial fraction of I--agents, $y_0 \to 0$. Figure
\ref{fig3} shows, in this case, the epidemics incidence as a
function of the recombination rate for several fixed infectivities.
We recall that, for $y_0 \to 0$, the mean field description predicts
a transcritical bifurcation between zero and positive incidence at a
critical infectivity $\lambda =1$, while in the absence of
recombination the incidence is identically zero for all
infectivities. Our numerical calculations show that, for
sufficiently large values of $r$, the transition is still present,
but the critical point depends on the recombination rate. As $r$
grows to infinity, the critical infectivity decreases approaching
unity.

Straightforward  linearization analysis of Eqs. (\ref{couples})
shows that  the state of zero incidence becomes unstable above the
critical infectivity
\begin{equation} \label{hc}
\lambda_c = \frac{r+1}{r-1}.
\end{equation}
This value is in excellent agreement with the numerical
determination of the transition point. Note also that Eq. (\ref{hc})
predicts a divergent critical infectivity for a recombination rate
$r=1$. This implies that, for $0\le r \le 1$, the transition is
absent and the disease has no incidence irrespectively of the
infectivity level. For $y_0 \to 0$, thus, the recombination rate
must overcome the critical value $r_c=1$ to find positive incidence
for sufficiently large infectivity. The critical line between zero
and positive incidence in the parameter plane of infectivity vs.
recombination rate, given by Eq. (\ref{hc}), is plotted in the
insert of the upper panel of Fig. \ref{fig2}.

\section{Conclusions}

We have studied the dynamics of SIR epidemics in a population where,
at any time, each agent forms a couple with exactly one neighbor,
but neighbors are randomly exchanged at a fixed rate. As it had
already been shown for the SIS epidemiological model
\cite{bouzat,vazquez}, this recombination of couples can, to some
degree, compensate the high disconnection  of the instantaneous
interaction pattern, and thus allow for the propagation of the
disease over a finite portion of the population. The interest of a
separate study of SIR epidemics is based on its peculiar dynamical
features: in contrast with SIS epidemics, it admits infinitely many
absorbing equilibrium states. As a consequence, the disease
incidence depends not only on the infectivity and the recombination
rate, but also on the initial fraction of infectious agents in the
population.

Due to the random nature of recombination, mean field--like
arguments provide exact equations for the evolution of couples
formed by agents in every possible epidemiological state. These
equations can be analytically studied in the limits of zero and
infinitely large recombination rates. The latter case, in
particular, coincides with the standard mean field description of
SIR epidemics.

Numerical solutions for intermediate recombination rates smoothly
interpolate between the two limits, except when the initial fraction
of infectious agents is vanishingly small. For this special
situation, if the recombination rate is below one recombination
event per couple per time unit (which equals the mean recovery
time), the disease does not propagate and its incidence is thus
equal to zero. Above that critical value, a transition appears as 
the disease infectivity changes: for small infectivities the
incidence is still zero, while it becomes positive for large
infectivities. The critical transition point shifts to lower
infectivities as the recombination rate grows.

It is worth mentioning that a similar transition between a state
with no disease and an endemic state with a permanent infection
level occurs in SIS epidemics with a vanishingly small fraction of
infectious agents \cite{bouzat,vazquez}. For this latter model,
however, the transition is present for any positive recombination
rate. For SIR epidemics, on the other hand, the recombination rate
must overcome a critical value for the disease to spread, even at
very large infectivities.

While both the (monogamous) structure and the (recombination)
dynamics of the interaction pattern considered here are too
artificial to play a role in the description of real systems, they
correspond to significant limits of more realistic situations.
First, the monogamous population represents the highest possible
lack of connectivity  in the interaction pattern (if isolated agents
are excluded). Second, random couple recombination preserves the
instantaneous structure of interactions and does not introduce
correlations between the individual epidemiological state of agents.
As was already demonstrated for SIS epidemics and chaotic
synchronization \cite{vazquez}, they have the additional advantage
of being analytically tractable to a large extent. Therefore, this
kind of assumption promises to become a useful tool in the study of
dynamical processes on evolving networks.

\begin{acknowledgements}
Financial support from SECTyP--UNCuyo and ANPCyT, Argentina, is
gratefully acknowledged.
\end{acknowledgements}

\end{document}